\documentclass[aps,superscriptaddress,twocolumn,showpacs,floatfix,notitlepage]{revtex4-2}

\usepackage{amsmath}
\usepackage[makeroom]{cancel}
\usepackage{amsfonts}
\usepackage{amssymb}
\usepackage{bm}
\usepackage{graphicx}
\usepackage{color}
\usepackage{xcolor}
\usepackage{xspace}
\usepackage[sort&compress]{natbib}
\usepackage{dsfont}
\usepackage{physics}
\usepackage{siunitx}
\usepackage{soul}
\usepackage{xfrac}
\usepackage[cal=boondox,scr=boondoxo]{mathalfa}
\usepackage{soul}
\usepackage{todonotes}

\DeclareFontFamily{OT1}{pzc}{}
\DeclareFontShape{OT1}{pzc}{m}{it}%
{<-> s * [1.15] pzcmi7t}{}
\DeclareMathAlphabet{\mathpzc}{OT1}{pzc}{m}{it}
\DeclareUnicodeCharacter{200D}{{\hskip 0pt}}

\newcommand{\ud}[1]{{#1^{\dagger}}}

\newcommand{\av}[1]{\langle  #1 \rangle}

\usepackage[colorlinks=true, linkcolor=blue, citecolor=magenta]{hyperref}

\allowdisplaybreaks

\begin{document}
\title{Unlocking multiphoton emission from a single-photon source \\through mean-field engineering}

\author{Sang Kyu Kim}
\email{sangkyu.kim@tum.de}
\affiliation{Walter Schottky Institut, TUM School of Computation, Information and Technology, and MCQST, Technische Universit\"at M\"unchen, 85748 Garching, Germany}%
\affiliation{Institute for Advanced Study, Technische Universit\"at M\"unchen, 85748 Garching, Germany}

\author{Eduardo Zubizarreta Casalengua}
\affiliation{Walter Schottky Institut, TUM School of Computation, Information and Technology, and MCQST, Technische Universit\"at M\"unchen, 85748 Garching, Germany}%

\author{Katarina Boos}
\affiliation{Walter Schottky Institut, TUM School of Computation, Information and Technology, and MCQST, Technische Universit\"at M\"unchen, 85748 Garching, Germany}

\author{\\Friedrich Sbresny}
\affiliation{Walter Schottky Institut, TUM School of Computation, Information and Technology, and MCQST, Technische Universit\"at M\"unchen, 85748 Garching, Germany}%

\author{Carolin Calcagno}
\affiliation{Walter Schottky Institut, TUM School of Computation, Information and Technology, and MCQST, Technische Universit\"at M\"unchen, 85748 Garching, Germany}%

\author{Hubert Riedl}
\affiliation{Walter Schottky Institut, TUM School of Natural Sciences, and MCQST, Technische Universit\"at M\"unchen, 85748 Garching, Germany}%

\author{Jonathan J. Finley}
\affiliation{Walter Schottky Institut, TUM School of Natural Sciences, and MCQST, Technische Universit\"at M\"unchen, 85748 Garching, Germany}

\author{Carlos~{Ant\'on-Solanas}}
\affiliation{Departamento de F\'isica de Materiales, Instituto Nicol\'as Cabrera,  Universidad Aut\'onoma de Madrid, 28049 Madrid, Spain}
\affiliation{Condensed Matter Physics Center (IFIMAC),Universidad Aut\'onoma de Madrid, 28049 Madrid, Spain}

\author{Fabrice P.~Laussy}
\affiliation{Instituto de Ciencia de Materiales de Madrid ICMM-CSIC, 28049 Madrid, Spain}

\author{Kai M\"uller}
\email{kai.mueller@tum.de}
\affiliation{Walter Schottky Institut, TUM School of Computation, Information and Technology, and MCQST, Technische Universit\"at M\"unchen, 85748 Garching, Germany}%

\author{Lukas Hanschke}
\affiliation{Walter Schottky Institut, TUM School of Computation, Information and Technology, and MCQST, Technische Universit\"at M\"unchen, 85748 Garching, Germany}%

\author{Elena {del Valle}}
\email{elena.delvalle.reboul@gmail.com}
\affiliation{Departamento de F\'isica Te\'orica de la Materia Condensada, Universidad Aut\'onoma de Madrid, 28049 Madrid, Spain}
\affiliation{Institute for Advanced Study, Technische Universit\"at M\"unchen, 85748 Garching, Germany}
\affiliation{Condensed Matter Physics Center (IFIMAC),Universidad Aut\'onoma de Madrid, 28049 Madrid, Spain}

\date{\today}

\begin{abstract}
  \bfseries 
Single-photon emission from a two-level system offers promising perspectives for the development of quantum technologies, where multiphotons are generally regarded as accidental, undesired and should be suppressed. In quantum mechanics, however, multiphoton emission can turn out to be even more fundamental and interesting than the single-photon emission, since in a coherently driven system, the multiphoton suppression arises from quantum interferences between virtual multiphoton fluctuations and the mean field in a Poisson superposition of all number states. Here, we demonstrate how one can control the multiphoton dynamics of a two-level system by disrupting these quantum interferences through a precise and independent homodyne control of the mean field. We show that, counterintuitively, quantum fluctuations always play a major qualitative role, even and in fact especially, when their quantitative contribution is vanishing as compared to that of the mean field.  Our findings provide new insights into the paradoxical character of quantum mechanics and open pathways for mean-field engineering as a tool for precision multiphoton control.
\end{abstract}
\maketitle

\begin{figure*}
  \includegraphics[width=1\linewidth]{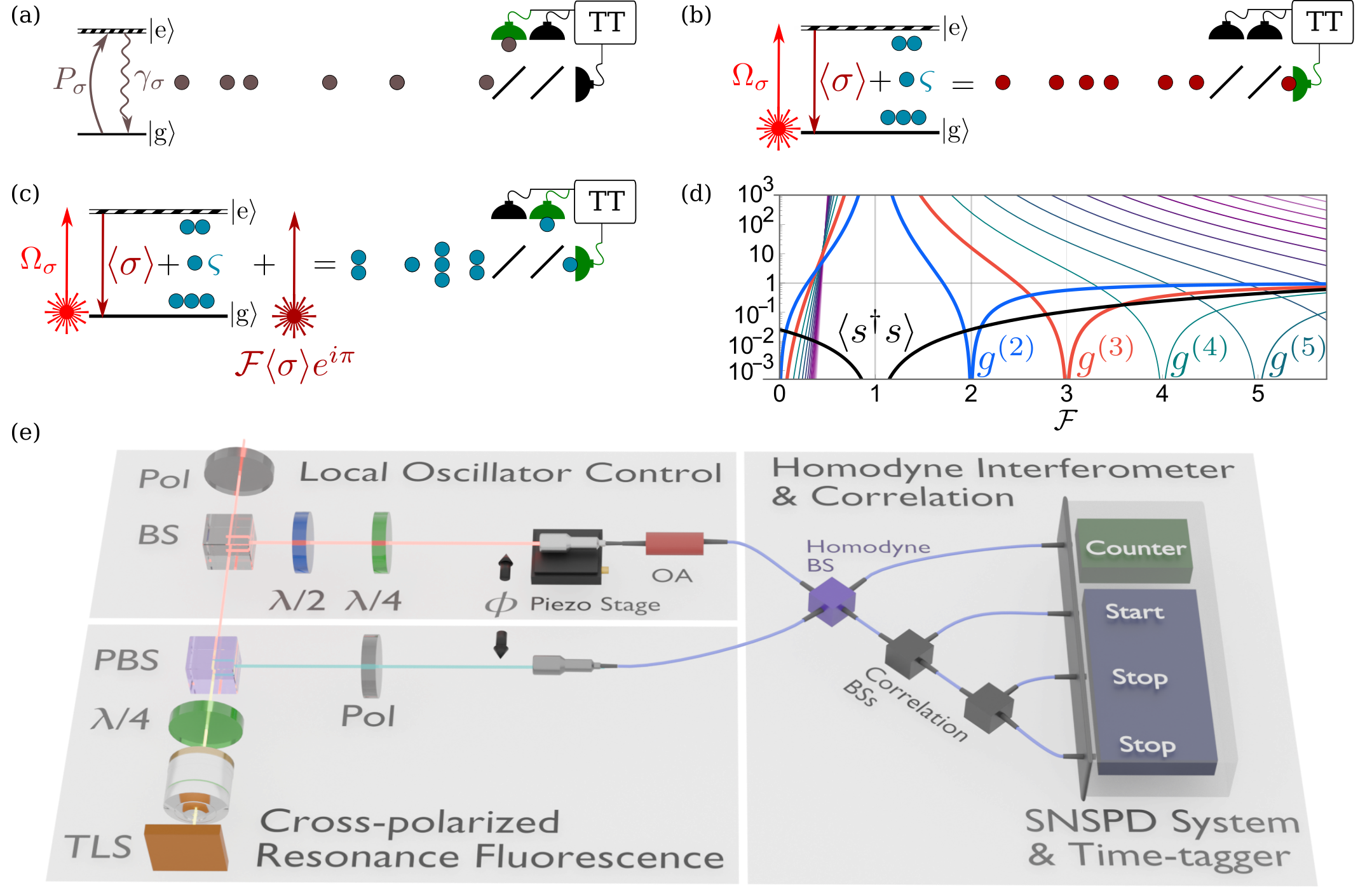}
  \caption{(a) For incoherent driving~$P_{\sigma}$, a TLS~$\sigma$ allows only one excitation at a time, leading to single-photon emission characterized by an extended Hanbury Brown-Twiss setup. TT, time-tagging unit. (b) Under coherent driving~$\Omega_\sigma$, however, the single-photon emission arises from interferences between the coherent mean field~$\langle\sigma\rangle$ and the quantum fluctuations~$\varsigma$. The probability amplitudes of multiphotons are canceled out by the destructive interferences with the mean field. (c) Disrupting the interferences by adding an external coherent field~$\mathcal{F}\langle\sigma\rangle e^{i\pi}$ can bring the multiphotons to light or suppress a given $n$-photon emission in an otherwise almost coherent field. (d) Glauber's~$g^{(n)}$ correlators provide suitable observables for the characterization of quantum light. In the Heitler regime, they all diverge at~$\mathcal{F}=1$ when the external field cancels~$\langle\sigma\rangle$---in which case multiphoton emission to all orders is observed from the TLS, beside its emission~$\langle s^\dagger s\rangle$ approaching~$0$. At~$\mathcal{F}=n$, each~$g^{(n)}$ is individually suppressed. (e) Experimental realization of the scheme by combining cross-polarized resonance fluorescence of a single quantum dot and a homodyne setup. An external coherent field---the local oscillator---is controlled in polarization, phase~$\phi$ and intensity~$\propto\mathcal{F}^2$. The homodyned signal is fed to a detection system, which allows us to study correlations with stabilized phase control. Pol, polarizer; BS, beam splitter; PBS, polarizing beam splitter; $\lambda/2$, half-waveplate; $\lambda/4$, quarter-waveplate; OA, optical attenuator; SNSPD, superconducting nanowire single-photon detector.}
\label{fig1}
\end{figure*}

Resonance fluorescence---the coherent excitation of a two-level system~(TLS) by a laser---has been a hallmark of the quantum theory of light-matter interaction since its early days~\cite{heitler_book54a}. This process has been extensively studied across various platforms, including atoms~\cite{jessen92a}, superconducting qubits~\cite{astafiev10a}, molecules~\cite{wrigge08a}, ions~\cite{hoffges97a}, chiral artificial atoms~\cite{joshi23a} and solid-state quantum dots~\cite{muller07a}, etc. The rich physics of resonance fluorescence has been investigated through a prolific literature over decades, ranging from multiphoton scattering~\cite{dalibard83a} to fluorescence from a squeezed vacuum~\cite{carmichael87a,toyli16a}, passing by interferences between past and future quantum states~\cite{campagneibarcq14a} as well as quantum dynamical~\cite{liu24a,boos24a} and nonlocal~\cite{lopezcarreno24a} aspects. Among these phenomena, the generation of single photons has received significant attention as a promising resource for quantum technologies.

In quantum mechanics, any observed outcome arises from the probability distribution over all possible states. Single-photon emission of a TLS under weak coherent driving can be thus described by a quantum superposition of all photon number states, where multiphotons remain virtual. By ``virtual'' we mean that their probability amplitudes play a role but cancel out in the measurement. Recently, the multiphoton aspect has been revisited for its spectacular manifestations of counterintuitive features of quantum physics. Prominently, Masters \emph{et al.} have shown how a single TLS can simultaneously emit two photons despite having only one transition available to do so~\cite{masters23a}. Liu \emph{et al.} have subsequently demonstrated entanglement of the emitted light and propelled its technological prospects~\cite{liu24a}. Manipulation of single or multiple photons has been achieved with more complex systems such as a TLS coupled to a cavity~\cite{faraon08a, flayac17b,snijders18a,vaneph18a, wang19a, tomm24a}, which relies on internal interference in the system but limits the tunability, and an ensemble of atoms mediating interference~\cite{prasad2020, corider2023}, by taking advantage of many-body enhancement. However, the multiphoton dynamics of the simplest system in quantum optics remains largely unexplored, leaving gaps in the understanding of the underlying physics.

In this work, we investigate the fundamental quantum interference of multiphoton fluctuations with a classical mean field in the resonance fluorescence of a TLS under weak driving. We observe that antibunching in all multiphoton correlations~(experimentally up to three-photon) turns into superbunching of all orders when the system is reduced to its quantum fluctuations. This result is achieved by disrupting the interferences with a full and independent control of the mean field. A precise admixture of the classical and quantum fields realizes individual suppression of photon numbers, revealing how the multiphoton coincidences behave independently from each other. This tunability opens up new possibilities for harnessing multiphoton physics.\\

The TLS, with its operator ~$\sigma\equiv\ketbra{g}{e}$, is the most fundamental quantum emitter. Its only transition from the excited~$\ket{e}$ to the ground~$\ket{g}$ state can be saturated, resulting in a stream of antibunched photons. While this paradigmatic description for single-photon emission from a TLS is accurate in cases such as with incoherent excitation~(Fig.~\ref{fig1}(a)), under coherent driving, an entirely different scenario arises~(Fig.~\ref{fig1}(b)). The coherent driving of an oscillator leads to a coherent response, even when this oscillator is quantum~\cite{allen_book87a}, and small nonlinearities can be dealt with perturbatively in the form of fluctuations. The fluctuations are obtained by subtracting the mean field~$\langle\sigma\rangle$ of the system from the TLS operator~$\sigma$:
\begin{equation}
  \label{eqn_def_varsigma}
  \varsigma\equiv\sigma-\langle\sigma\rangle\,,
\end{equation}
where $\varsigma$ is the quantum fluctuation operator. When the driving~$\Omega_\sigma$ is weak compared to the radiative decay rate~$\gamma_\sigma$ of the emitter, the ratio between the intensity of the mean field~$|\langle\sigma\rangle|^2$ and the intensity of the fluctuations~$\langle\ud{\varsigma}\varsigma\rangle$ can be made arbitrarily large:
\begin{equation}
  \label{eq:Mon16Sep101459CEST2024}
  \langle\ud{\varsigma}\varsigma\rangle\ll|\langle\sigma\rangle|^2 \,.  
\end{equation}
In this so-called Heitler regime~\cite{heitler_book54a}, i.e., when~$\Omega\ll 1$, where~$\Omega\equiv\Omega_\sigma/\gamma_\sigma$, the total intensity~$\langle\ud{\sigma}\sigma\rangle=|\langle\sigma\rangle|^2+\langle\ud{\varsigma}\varsigma\rangle$ is composed of~\cite{meystre2007a}
\begin{equation}
  \label{eqn_intensities}
  |\langle\sigma\rangle|^2=\left({2\Omega\over1+8\Omega^2}\right)^2
  \quad\text{and}\quad
  \langle\ud{\varsigma}\varsigma\rangle={8 \Omega^2 |\langle\sigma\rangle|^2} \,.
\end{equation}
In a classical setting, when fluctuations are significantly smaller than the mean field, they are regarded as negligible or treated as perturbative corrections. In a fully quantum description, this perturbative picture breaks down regardless of the quantitative imbalance, and what occurs is a much more dramatic excitation of the quantum field into a superposition of all its possible multiphoton states. This outcome is surprising as one might expect that in this weak driving regime, multiphotons would play no role whatsoever. 
As far as single-photon emission is concerned, interferences of the quantum fluctuations with the classical mean field result in the finally observed emission of single photons, out of a dormant quantum field simmering with multiphotons. These multiphotons can be revealed by disrupting their interferences. Experimentally, this can be achieved by controlling the mean field---a coherent component---through the so-called homodyne technique~\cite{vogel91a}. It consists in admixing a local oscillator~(LO) with the signal to precisely adjust the coherent fraction. This technique proved extremely effective to observe small quantum effects obscured by a strong classical field~\cite{breitenbach97a,schulte15a,fischer16a}. Here, one can seize control of the coherent field to unknit the single-photon emission into multiphotons~(Fig.~\ref{fig1}(c)). The external LO field is represented as a coherent state
\begin{equation}
  \label{eq:Mon16Sep120257CEST2024}
  \ket{\mathcal{F}\langle\sigma\rangle e^{i\phi}} \,,
\end{equation}
whose amplitude we write as a factor~$\mathcal{F}$ of~$\langle\sigma\rangle$. The relative phase~$\phi=\pi$ is set to be opposite to the phase of the mean field. As a result, the intensity of the total signal~$s\equiv\sigma+\mathcal{F}\langle\sigma\rangle e^{i\pi}$ which is, in general,
\begin{equation}
  \label{eqn_intensity_signal}
  \langle\ud{s}s\rangle=(\mathcal{F}-1)^2|\langle\sigma\rangle|^2+\langle\ud{\varsigma}\varsigma\rangle \,,
\end{equation}
reduces to the quantum fluctuations~$\langle\ud{\varsigma}\varsigma\rangle$ when~$\mathcal{F} = 1$. 

Next, we show how the small fluctuations govern the multiphoton physics of the system. The best way to characterize quantum light is through the standard observables in quantum optics: Glauber's $n$th-order correlation functions~$g^{(n)}(\tau_1,\dots,\tau_{n-1})$. They quantify the density of $n$-photon detections separated by times~$\tau_i$. The joint detection of~$n$ photons, i.e., with~$\tau_i=0$ for all~$i$, measures by how much a coincidence is magnified~(or suppressed if~$<1$) as compared to an uncorrelated signal of same intensity. For the problem at hand of unleashing multiphoton emission from a TLS by admixing an external LO field, the $n$-photon coincidences can be obtained exactly for any driving~(Methods). In the Heitler limit of~$\Omega\to0$, they read
\begin{equation}
  \label{eqn_gn}
  g^{(n)}(0)={\mathcal{F}^{2(n-1)}(\mathcal{F}-n)^2\over(\mathcal{F}-1)^{2n}} \,.
\end{equation}
From the denominator, one can see that these multiphoton observables, shown in Fig.~\ref{fig1}(d), diverge for all~$n$ when~$\mathcal{F}=1$. They are superbunched for all photon numbers according to Eq.~(\ref{eqn_gn}). This so-called unconventional bunching~\cite{zubizarretacasalengua20a} occurs when the mean field~$\langle\sigma\rangle$ is canceled completely from the signal, leaving only the quantum fluctuations~$\varsigma$. It is remarkable that subtracting the mean field from a TLS leads to strong multiphoton emission to all orders.

Now turning to the numerator of Eq.~(\ref{eqn_gn}), one can see that the multiphoton correlations have two zeros: the first one for all~$n$ at~$\mathcal{F}=0$, i.e., without an external field. This case is the standard resonance fluorescence, which exhibits both the conventional antibunching~\cite{zubizarretacasalengua20a}---single-photon emission---and sub-natural linewidth~\cite{nguyen11a,matthiesen12a}, though these two properties cannot be observed simultaneously~\cite{lopezcarreno18b,phillips20a,hanschke20a,zubizarretacasalengua24b}.

More striking features occur with the second zero in Eq.~(\ref{eqn_gn}) which is~$n$-dependent with~$g^{(n)}(0)\to0$ at~$\mathcal{F}=n$. Although~$\mathcal{F}$ is a continuous variable, as befits a classical field, it triggers a strong response of the system when taking integer values, which is a manifestation of the interplay between interfering continuous and quantized fields. Unlike the previous case with~$\mathcal{F}=0$, these multiphoton resonances are not degenerate. One can suppress any given photon number individually without strongly affecting the other correlators, realizing the unconventional antibunching~\cite{zubizarretacasalengua20a}. In particular, one can suppress two-photon emission only, i.e., $g^{(2)}(0)\to 0$ at~$\mathcal{F}=2$. In this case, due to the proximity to the divergence at~$\mathcal{F}=1$, all the other photon-number coincidences remain much larger than would be expected on accounts of random events alone, with~$g^{(n)}(0)\gg 1$ for~$n\ge 3$. In other words, the suppression of two-photon coincidences does not preclude increased coincidences of higher numbers of photons such that
\begin{equation}
  \label{eqn_F2}
  g^{(2)}(0)\ll g^{(3)}(0) \,.
\end{equation}
At~$\mathcal{F}=3$, one can realize a suppression of the three-photon coincidences without strong suppression of the two-photon ones,~$g^{(3)}(0)\to 0$ while~$g^{(2)}(0)\lessapprox 1$ and reverse the trend of Eq.~(\ref{eqn_F2}), i.e., qualitatively
\begin{equation}
  \label{eqn_F3}
  g^{(2)}(0)\gg g^{(3)}(0) \,.
\end{equation}

\begin{figure}[h]
    \includegraphics[width=\linewidth]{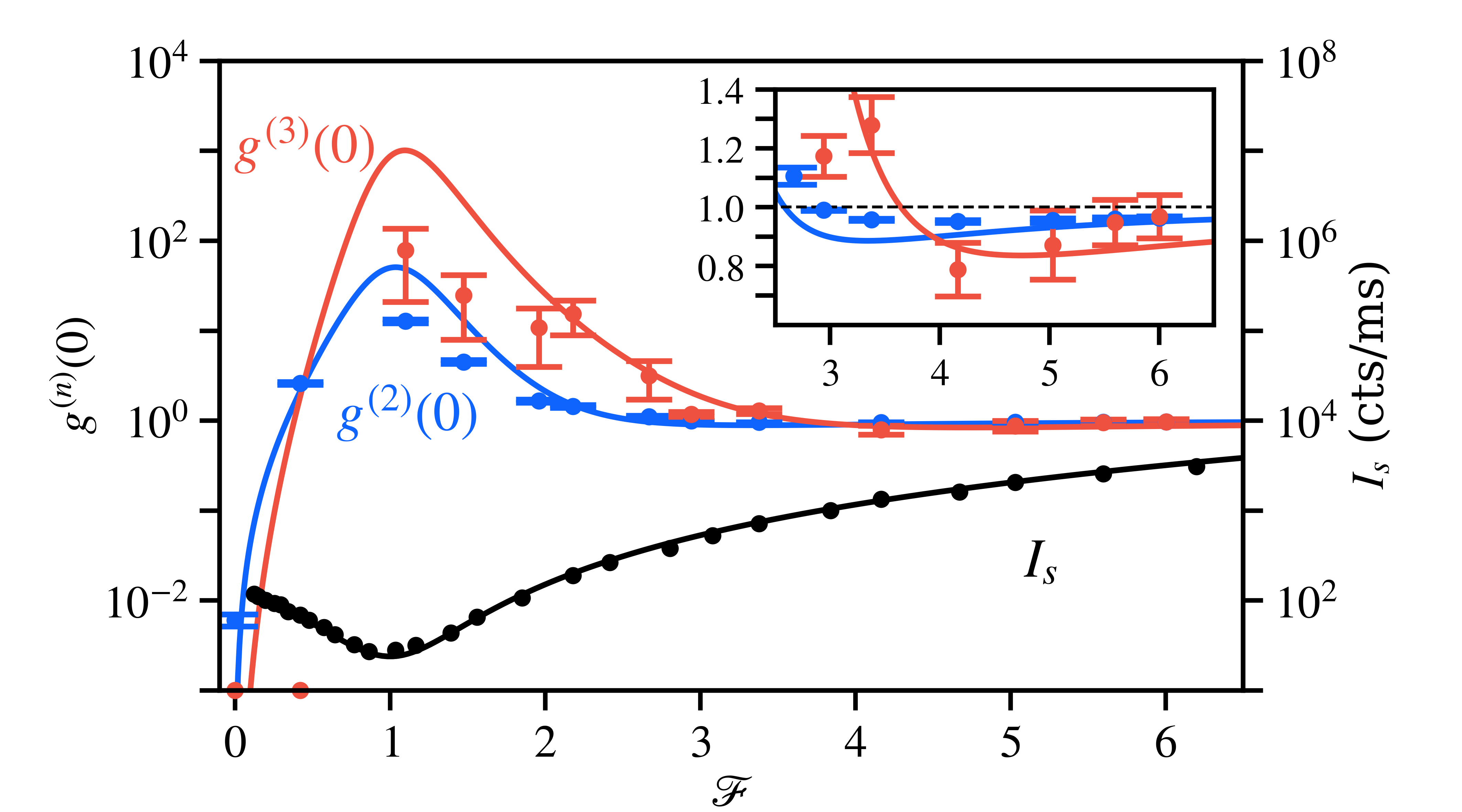}
    \caption{Experimental results~(symbols) and theoretical predictions~(solid lines) for intensity~(black), along with~$g^{(2)}(0)$~(blue) and~$g^{(3)}(0)$~(red) of the homodyned signal, as functions of the LO field amplitude~$\mathcal{F}$ under a driving~$\Omega\approx0.15$. At~$\mathcal{F}=0$, without admixing with an external field, multiphoton events remain virtual due to destructive interferences of their probability amplitudes, resulting in the emission of single photons. At~$\mathcal{F}=1$, the external field cancels the mean field of the system and thus lays bare its quantum fluctuations. The multiphoton nature of the fluctuations is revealed in strong~$g^{(2)}(0)$ and~$g^{(3)}(0)$ superbunching despite a drop of the intensity by about an order of magnitude. For a sizable coherent field, one can suppress $n$-photon emission independently. The cases~$n=2$ and~$n=3$ are shown and magnified in the inset, confirming the transition from two-photon suppression to three-photon suppression. The two undetermined cases with~$g^{(3)}(0)=0$, arising from zero coincidence events throughout the entire integration time, are shown on the horizontal axis.}
    \label{fig2}
\end{figure}

In our experiment, the TLS is realized with a single InGaAs quantum dot under weak driving~(Methods). Our setup with multiphoton coincidence counting units to characterize the homodyned signal is sketched in Fig.~\ref{fig1}(e). In Fig.~\ref{fig2}, we study the intensity and multiphoton coincidences as the mean field is manipulated. Experimentally, the measured intensity~$I_{s}$ of the admixture is proportional to the theoretical quantity~$\langle\ud{s}s\rangle$, scaled by overall experimental efficiency, emitter decay rate and other relevant factors. The intensity as a function of the external field~$\mathcal{F}$~(black symbols) allows us to access the intensity of the coherent field,~$I_{\langle\sigma\rangle}\propto|\langle\sigma\rangle|^2$, and that of the quantum field,~$I_{\varsigma}\propto\langle\varsigma^\dagger\varsigma\rangle$, as well as to extract the applied driving strength experienced by the system. We obtain~$I_{\langle\sigma\rangle}\approx\SI{252.2}{cts/\ms}$ and~$I_{\varsigma}\approx\SI{47.5}{cts/\ms}$ at the driving~$\Omega\approx0.15$~(Supplementary Information). For the given driving, the contribution of the coherent mean field to the total emission is about~$\SI{84}{\percent}$, satisfying Eq.~(\ref{eq:Mon16Sep101459CEST2024}).
As~$\mathcal{F}$ increases from 0 to 1, the intensity of the admixture decreases by roughly an order of magnitude as the coherent emission is gradually removed. When only the fluctuations remain in the signal, multiphotons emerge with strong bunching correlations. Theoretically, we observe~$g^{(n+1)}(0) \gg g^{(n)}(0) \gg 1$ for all~$n\geq2$, and experimentally, we confirm~$g^{(3)}(0) \gg g^{(2)}(0) \gg 1$ at~$\mathcal{F}=1$ in Fig.~\ref{fig2}. While this quantum effect becomes more pronounced as the driving decreases, in the laboratory, the Heitler limit is an asymptotic ideal which must be compounded with experimental limitations such as efficiency and stability. For finite driving, we observe that the correlations retain the same structure but become smoother, decrease in contrast, and shift in position.

The non-degenerate multiphoton antibunching resonances are observed with up to three photons in the inset of Fig.~\ref{fig2}. These resonances at the finite driving are not as prominent as in the mathematical limit. However, all the qualitative relationships of Eqs.~(\ref{eqn_F2}) and~(\ref{eqn_F3}) are satisfied and in good agreement with the theoretical prediction for this driving. We observe two-photon antibunching and three-photon bunching at~$\mathcal{F}=3.38$. Namely, with~$g^{(2)}(0)=0.957\pm0.004$ and~$g^{(3)}(0)=1.28\pm0.10$, we indeed have~$g^{(2)}(0)<1<g^{(3)}(0)$. At~$\mathcal{F}=4.17$, we confirm the opposite stronger three-photon suppression with~$g^{(2)}(0)=0.951\pm0.004$ and~$g^{(3)}(0)=0.79\pm0.09$. Consequently, we have demonstrated that multiphoton coincidences can be enhanced or suppressed, together or independently, depending on how the mean field ties them together.\\

\begin{figure*}
  \includegraphics[width=\linewidth]{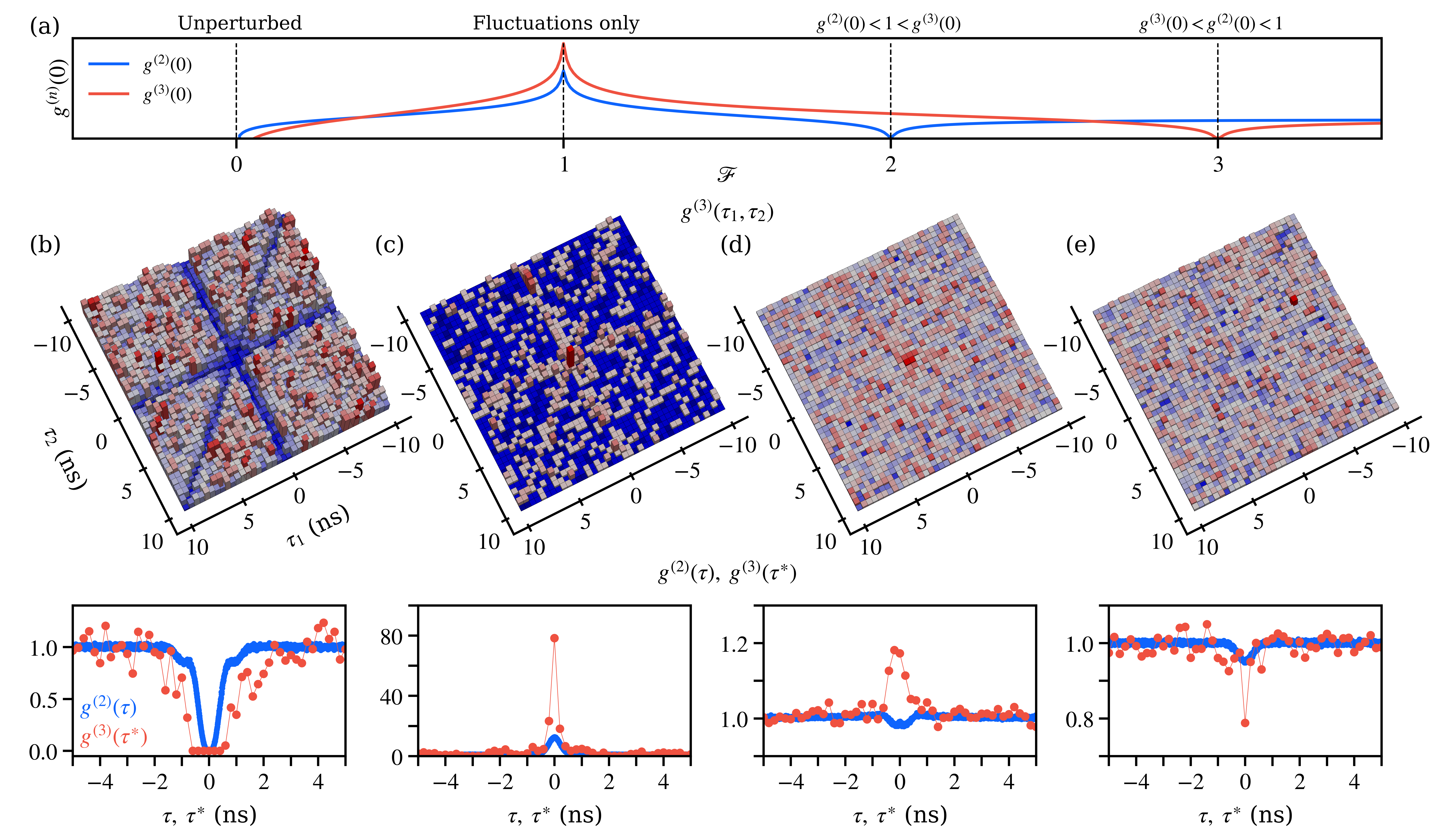}
  \caption{(a) Calculated multiphoton observables in the Heitler
    limit~$\Omega\to0$. (b--e) For~$\Omega\approx0.15$ normalized third-order coincidences~$g^{(3)}(\tau_1,\tau_2)$ with time
    delays~$\tau_1$ and~$\tau_2$~(second row) and their integrated
    results~$g^{(3)}(\tau^*)$ with second-order correlation~$g^{(2)}(\tau)$~(third
    row) for four configurations of interest. (b) With no external
    field, $\mathcal{F}=0$,  all multiphoton probabilities are suppressed, producing single-photon emission. In~$g^{(3)}(\tau_1,\tau_2$), this is
    visible as three depleted lines~($g^{(2)}(0)\ll 1$) and a wide dip at
    their intersection~($g^{(3)}(0,0)\ll 1$) in an otherwise uncorrelated
    background, satisfying~$g^{(3)}(0)\ll g^{(2)}(0)\ll 1$. (c)
    Canceling the mean field, with~$\mathcal{F}=1$, we observe
    strong multiphoton bunching from the quantum fluctuations, visible
    as three lines of the signal standing on the vacuum. This realizes~$g^{(3)}(0)\gg g^{(2)}(0)\gg 1$. (d)
    Changing the sign of the mean field realizes the
    counterintuitive relation~$g^{(2)}(0)<1<g^{(3)}(0)$. Two-photon emission is
    suppressed, but not higher-order emissions, thus disqualifying
    this regime as a single-photon source. (e) Individual suppression of the three-photon component. Further increasing the mean field realizes the next case where~$g^{(3)}(0) < 1$ while
    ~$g^{(2)}(0)\lessapprox 1$. }
    \label{fig3}
\end{figure*}

To discuss the multiphoton physics underlying these counterintuitive quantum phenomena in more detail, we examine four distinct regimes with a theoretical overview~(Fig.~\ref{fig3}(a)) across continuous variations of~$\mathcal{F}$ in the Heitler limit. Experimental results of~$g^{(3)}(\tau_1,\tau_2)$, where~$\tau_1$ and~$\tau_2$ are the time differences of events detected by the second and third detectors relative to the first, are shown in Fig.~\ref{fig3}(b)-(e) under a finite driving~$\Omega\approx0.15$, along with~$g^{(2)}(\tau)$ and radially integrated~$g^{(3)}(\tau^*)$, where~$\tau^*$ represents the effective time delay for three-photon events as~$|\tau^*|=\sqrt{\tau_1^2+\tau_2^2}$~(Methods).

Our observations are unequivocal: The leftmost column describes the usual Heitler regime of resonance fluorescence. At~$\mathcal{F}=0$, while the emission is predominantly coherent, it is antibunched in all multiphoton correlators. The antibunching in the second-order,~$g^{(2)}(0) = 0.006\pm0.001$, is also seen in~$g^{(3)}(\tau_1, \tau_2)$ of Fig.~\ref{fig3}(b) as three lines with vanishing coincidences across the whole landscape, when two of the three detectors are triggered at the same time, i.e.,~$\tau_1 = 0$, $\tau_2 = 0$ or~$\tau_1=\tau_2$. At their intersection,~$\tau_1=\tau_2=0$, no photon triplet within~\SI{1.2}{\ns} time delay~(corresponding to the center plateau of the integrated~$g^{(3)}(\tau^*)$) was observed over the entire integration time~($>19$ hours). For an uncorrelated signal of the same intensity, approximately~12 three-photon coincidence events would be expected from the averaged uncorrelated three-photon events~$\bar{G}^{(3)}(\infty)=2.02$ for~$\SI{200}{\ps}$ binning.

The second column, Fig.~\ref{fig3}(c), describes the case at~$\mathcal{F}\approx1$, showcasing our homodyne technique that gives us access to the naked quantum fluctuations~$\varsigma$. How the quantum signal looks like in~$g^{(3)}(\tau_1,\tau_2)$ landscape is striking: the considerable drop in classical emission but the persistence of simultaneous two- and three-photon emission produces bunched diagonals that stand on the vacuum. Although the three-photon coincidence~$g^{(3)}(0,0)$, as a third-order process, is measured from a much more scarce signal than~$g^{(2)}(0)$, its much stronger deviation from the classical case makes its features more discernible. We obtain~$g^{(2)}(0)=12.7\pm0.6$ and~$g^{(3)}(0)=78\pm57$, confirming the multiphoton nature of the fluctuations. 
Our findings can be seen directly from the photon number probability distribution, which can be extracted from the Glauber correlators~\cite{zubizarretacasalengua17a}. The probability of~$n$-photon coincidences~$p(n)$ of the homodyned signal is given by the Poissonian distribution of the external LO modulated by the quantum fluctuations~(Methods). At~$\mathcal{F}=1$, the TLS could be seen as a single-photon filter of the external LO, suppressing the single-photon emission in the admixture, while the probability of two-photon coincidences~$p(2)$ is identical to that of the LO as discussed earlier~\cite{wang19a,tomm2023, prasad2020}. However, this filter picture does not hold for higher photon numbers. Instead, the weak signal displays~$(n-1)^2$ times more~$n$-photon coincidences than the external field does. This shows that, from three photons upward, a coherent state can extract more multiphoton events from the quantum fluctuations than it has itself. Thus, there exists a multiphoton amplification for~$n\geq3$. Importantly, the one-photon probability still remains dominant among all~$n$-photon probabilities even at its strongest suppression at~$\mathcal{F}=1$~\cite{zubizarretacasalengua25a}.  

The third and fourth columns, where the mean field is again sizable, compellingly show the qualitative~(but not yet quantitative) relationships~(\ref{eqn_F2}) and~(\ref{eqn_F3}), which complete the versatility of the interplay between the mean field and the fluctuations. Bunching and antibunching of photon triplets are directly visible at the center of the three-photon landscapes~
$g^{(3)}(\tau_1,\tau_2)$ in Fig.~\ref{fig3}(d) and (e), respectively. By introducing more external field  as the LO amplitude~$\mathcal{F}$ increases from~2.94 to~4.17, the antibunching resonance~$g^{(n)}(0)<1$ shifts from~$n=2$ to~$n=3$ as shown by the transition of~$g^{(3)}(0)$ from~$1.17\pm0.07$ to~$0.79\pm0.09$ while~$g^{(2)}(0)$ remains below~$1$. In~Fig.~\ref{fig3}(d), one can see this independent behavior of the multiphoton observables that suppresses two-photon events, while three-photon ones become more likely than an uncorrelated signal with same intensity. With further increase of the external field, three-photon events are the most strongly suppressed among all~$n$. This is counterintuitive, since one would expect the impact of the fluctuations to become negligible due to a strong external field at~$\mathcal{F}\gg1$. However, instead of smothering the quantum attributes by making the fluctuations even more negligible, the classical field singles out strong~$g^{(n)}(0)\ll 1$ for high-order~$n\geq2$ quantum resonances, with no theoretical upper limit neither on~$n$ nor on the total field intensity. In this regime, a strong destructive quantum interferences occurs selectively for a specific photon number that is determined by the amplitude of the external field.\\

\begin{figure}
    \includegraphics[width=\linewidth]{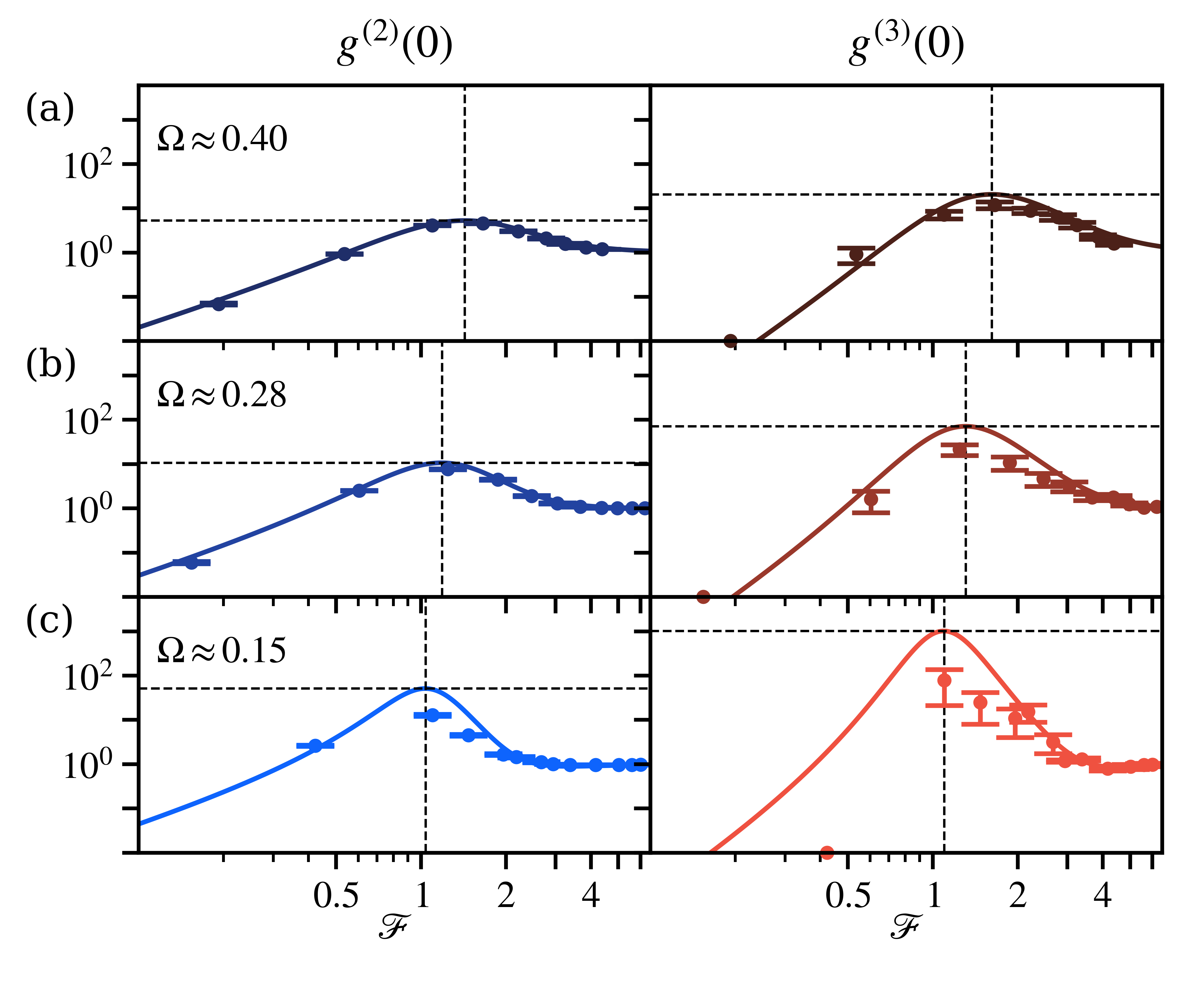}
    \caption{Power-dependent correlations of~$g^{(2)}(0)$~(left) and ~$g^{(3)}(0)$~(right) with decreasing driving of (a)~$\Omega\approx0.40$, (b)~$\Omega\approx0.28$ and (c)~$\Omega\approx0.15$. The multiphoton correlations of the fluctuations at~$\mathcal{F}\approx1$ increases as driving decreases, i.e., as the system approaches the Heitler limit, also with resonances converging to each other and towards~$\mathcal{F}=1$. Theoretical and experimental results are given as solid lines
      and symbols, respectively, dashed guidelines represent the strongest bunching resonances. All the
      data presented in the main text is for the lowest driving (c) where the correlations exhibit the strongest variation. As a result
      of setup instability over the long acquisition time required due to the weak signal, deviations from the theory are also the largest for this driving strength. The undetermined cases with~$g^{(3)}(0)=0$, arising from zero coincidence events throughout the entire integration time, are shown on the horizontal axis.}
      \label{fig4}
\end{figure}

In Fig.~\ref{fig4}, we provide evidence for such non-classical observations being enhanced in the deeper Heitler regime. The correlations at three driving strengths~($\Omega\approx0.40$,~$0.28$ and~$0.15$) show that the multiphoton effects are indeed magnified as the system approaches the lower driving case, even though the intensity contribution of the quantum fluctuations decreases~($\langle\ud{\varsigma}\varsigma\rangle/\langle\ud{\sigma}\sigma\rangle$ of~$\SI{56}{\percent}$,~$\SI{38}{\percent}$ and~$\SI{16}{\percent}$, Supplementary Information). Furthermore, the two- and three-photon resonances converge toward the common theoretical value~$\mathcal{F}=1$. The largest deviation between the measured values and the theoretical model is found with the smallest driving, which we primarily attribute to the limited setup stability during the long integration time of 277 hours for~$\Omega\approx0.15$. The larger signal allows us to conduct the measurements over smaller periods of time~(25.2 hours for~$\Omega\approx 0.28$ and 4.5~hours for~$\Omega\approx 0.40$), thereby mitigating this limitation. Consequently, there is a closer agreement with the theoretical model at higher powers, where quantum effects are, however, attenuated. One could, with brighter emitters as well as more efficient and stable setups, optimize the features we have reported by further reducing~$\Omega$, i.e., accessing deeper the Heitler regime.\\

In conclusion, we have experimentally demonstrated the external control of the quantum emission from a TLS under weak coherent driving. This control unveils the intrinsic multiphoton nature of the paradigmatic, fundamental, and simplest quantum optical emitter, producing both antibunching and bunching across all orders as well as their independent suppression. The multiphoton physics of the system is governed by quantum interferences of multiphoton fluctuations with a coherent mean field, which we manipulate independently. These quantum effects become increasingly pronounced as the system approaches the deeper Heitler regime. This multiphoton aspect similarly applies to quantum emitters in cavities, coupled TLSs and strongly-correlated phases in condensed matter systems. Our results open the door to further exploit such effects as a resource, for example, feeding a deterministic photon sorting system~\cite{yang22a} for photonic phase-transistors that take advantage of the multiphoton amplification that we have identified for~$n\geq3$, or through the realization of a simultaneous sub-natural linewidth single-photon source~\cite{lopezcarreno18b}. Building on our demonstration, we anticipate that mean-field control will serve as an universal key for unlocking multiphoton emission in coherently driven quantum systems and for enabling the generation of non-classical light beyond single photons.

\section*{Methods}
\paragraph*{\bfseries Theoretical model for multiphoton observables.} 
A TLS under coherent driving at resonance is modeled by the Hamiltonian, $H = \Omega_\sigma(\sigma^\dagger+\sigma)$. Solving the master equation,~$\partial_t{\rho} = i[\rho, H]+(\gamma_\sigma/2)\mathcal{L}_\sigma\rho$, in the Lindblad form, where the superoperator~$\mathcal{L}_\sigma\rho =
2 \sigma \rho \ud{\sigma} - \ud{\sigma} \sigma \rho - \rho \ud{\sigma} \sigma$, yields the following solution for the steady state
\begin{equation}
    \rho = \begin{pmatrix} 1-\langle n_\sigma\rangle & \av{\sigma}^* \\ \av{\sigma} & \langle n_\sigma\rangle\end{pmatrix} \,,
\end{equation}
where
\begin{equation}
    \langle n_\sigma\rangle  = \frac{4 \Omega_\sigma^2}{\gamma_\sigma^2 + 8 \Omega_\sigma^2}\
    \quad\text{and}\quad
    \langle \sigma\rangle  = \frac{2 i \Omega_\sigma \gamma_\sigma}{\gamma_\sigma^2 + 8 \Omega_\sigma^2} 
\end{equation}
are the population and the mean field of the system, respectively.
The correlation functions of the homodyned signal can be derived by substituting~$\sigma$ with~$s = \sigma + \mathcal{F} \av{\sigma} e^{i \phi} $~\cite{zubizarretacasalengua20b}:
\begin{multline}
\label{eq:homGns}
   G^{(n)}(0) =  \av{s^{\dagger n} s^n} = \left(\mathcal{F}^2 |\av{\sigma}|^2 \right)^n (\mathcal{F} \gamma_\sigma)^{-2} \cross \\
    [(n^2 + \mathcal{F}^2) \gamma_\sigma^2 + 8 n^2 \Omega_\sigma^2 + 2 n \mathcal{F} \gamma_\sigma^2 \cos(\phi)] \,,
\end{multline}
from which the Glauber's correlation functions read
\begin{multline}
   g^{(n)}(0) =  \frac{\av{s^{\dagger n} s^n}}{\av{s^\dagger s}^n} =  \\
   \frac{(\mathcal{F} \gamma_\sigma )^{2(n-1)} [(n^2 + \mathcal{F}^2) \gamma_\sigma^2 + 8 n^2 \Omega_\sigma^2 + 2 n \mathcal{F} \gamma_\sigma^2 \cos(\phi)]}{
[\mathcal{F}^2 \gamma_\sigma^2 + 2 \mathcal{F} \gamma_\sigma^2 \cos (\phi)+ (\gamma_\sigma^2 + 8 \Omega_\sigma^2)]^n   } \,.
\end{multline}
This result simplifies to Eq.~(\ref{eqn_gn}) for vanishing driving~($\Omega_\sigma / \gamma_\sigma \rightarrow 0$) and out-of-phase~($\phi=\pi$) condition.

The photon number probability distribution~$p(n)$ of the signal, representing the diagonal elements of the effective quantum state, accounts for the correlations of the admixture and can be reconstructed from the Glauber's correlators as~\cite{zubizarretacasalengua17a}
\begin{equation}
  p(n)= \sum_{k = 0}^\infty  \frac{(-1)^k}{k!} \frac{G^{(n+k)} (0)}{n!} \,.
\end{equation}
Substituting~$G^{(n} (0)$ in this formula by Eq.~(\ref{eq:homGns}), the analytical solution of the photon number distribution can be obtained regardless of the driving strength. 
However, for simplicity, we consider the limiting case of weak driving:
\begin{equation}
  \label{eqn_pn}
  p(n)=p_\mathrm{coh}(n)\mathcal{M}_\mathcal{F}(n) \,,
\end{equation}
where~$p_\mathrm{coh}(n)=\braket{n}{\mathcal{F}\langle\sigma\rangle e^{i\phi}}$ is the Poisson distribution of the external coherent
state as shown in Eq.~(\ref{eq:Mon16Sep120257CEST2024}). $\mathcal{M}_\mathcal{F}(n)$ is a multiphoton quantum correction that modulates this coherent state:
\begin{multline}
  \label{eqn_Mf}
  \mathcal{M}_\mathcal{F}(n)\equiv\\\left(1-{n\over\mathcal{F}}\right)^2-\left(1+2n-2\mathcal{F}-{2n^2\over\mathcal{F}^2}\right)\left({2\Omega}\right)^2+o(\Omega^4) \,.
\end{multline}
This description of the quantum interference predisposes one to see the TLS as a photon-number-sensitive filter, but as discussed in the text, this is in fact more like a multiphoton amplifier~(from three-photon upward) since more multiphotons can be emitted than are available anywhere in the system~\cite{zubizarretacasalengua25a}.\\

\paragraph*{\bfseries Quantum dot as a TLS.}
The experiments are performed with a single self-assembled quantum dot, where a neutral exciton exhibits a fine-structure splitting of~$\SI{790}{\MHz}$. One dipole of the neutral exciton transitions, with an emission wavelength of~$\SI{910}{\nm}$ and a radiative lifetime of~$\SI{216}{\ps}$, is driven by a continuous-wave laser. The narrow laser's linewidth~($\SI{<10}{\kHz}$) allows for resonant driving of the single dipole in our quantum dot, thereby realizing a TLS. The sample is cooled down to $\SI{4.2}{K}$ in a helium dip stick. A Schottky diode structure embedding the studied quantum dot allows us to stabilize the electronic environment and to fine-tune the emission wavelength of the system via the quantum-confined Stark effect~\cite{warburton2000}. Below the quantum dot layer, a distributed Bragg reflector enhances the collection of resonance fluorescence from the system. The reflection of the excitation laser is effectively suppressed by cross-polarized excitation and detection~\cite{vamivakas2009}. \\

\paragraph*{\bfseries Homodyne mean-field engineering.}
As shown in Fig.~\ref{fig1}(e), our homodyne interferometer is constructed similarly to Ref.~\cite{schulte15a}, however, instead of using the reflection of the excitation as a LO input, we pick off the excitation beam before interaction with the sample. This ensures that the LO remains a coherent state, free from any emission by the system. The polarization of the LO is carefully controlled by a set of motorized half- and quarter-waveplates to align with the polarization of the collected system emission. We control the intensity of the LO, which is proportional to~$\mathcal{F}^2$, using a fiber-based optical attenuator. The phase difference~$\phi$ between the emission and the LO is either scanned or stabilized upon measurement requirements, by adjusting  the propagation path length of the LO with a linear translation stage. The stabilization during correlation measurements is managed by a software PID control. In each control cycle~($\SI{10}{\ms}$), interference visibility is calculated from the detected counts of all channels, and the stage position is adjusted to maintain a given target visibility. For passive stabilization, we place our setup in a home-made stabilization box whose interior is covered with acoustic absorption foam to reduce the impact of environmental vibrations. By interfering the LO and the system emission in a polarization-maintaining fiber-based beam splitter, we minimize possible degradation of signal such as polarization rotation and mode mismatch caused from environmental fluctuations, e.g. temperature drift and mechanical vibrations. \\

\paragraph*{\bfseries Three-photon correlation.} 
To examine the second- and third-order correlations, one output of the homodyne beam splitter in Fig.~\ref{fig1}(e), where the signal of interest is desired, is split by two cascaded 50:50 fiber-based beam splitters and collected by three superconducting nanowire single-photon detectors, implementing an extended Hanbury Brown-Twiss setup. Time-tags of detection events are recorded by a multichannel time-tagger with an overall timing jitter of approximately~$\SI{20}{\ps}$. This configuration allows for the measurement of the unnormalized three-photon correlation~$G^{(3)}(\tau_1, \tau_2)$ through a two-dimensional histogram. Such time-correlated multiphoton measurements are technically challenging, since we operate with a single TLS, which we furthermore push in a regime of vanishing classical emission. A large bin size in these correlation measurements can mitigate limited signal counts, but it also reduces the timing resolution. As a result, features like bunching or antibunching become less noticeable, similarly to the effects of detector timing jitter~\cite{lopezcarreno2022a}. A bin size of~$\SI{500}{\ps}$ is chosen for the~$g^{(3)}(\tau_1,\tau_2)$ results presented in Fig.~\ref{fig2}, achieving visible features with a reasonable signal-to-noise ratio. We maximize the visibility of the signal and further reduce the bin size by performing a radial integration along the antidiagonal line~($\tau_1 = -\tau_2$) in ~$G^{(3)}(\tau_1, \tau_2)$ and obtain~$g^{(3)}(\tau^*)$ by normalizing the integrated~$G^{(3)}(\tau^*)$ with the averaged uncorrelated events~(Supplementary Information). As a result, the final values are extracted with a bin size of~$\SI{200}{\ps}$ for~$g^{(3)}(\tau^*)$.\\

\section*{Acknowledgements}
We gratefully acknowledge financial support from the German Federal Ministry of Education and Research via the funding program Photonics Research Germany (Contract No. 13N14846) and the Deutsche Forschungsgemeinschaft (DFG, German Research Foundation) via projects MU 4215/4-1 (CNLG), INST 95/1220-1 (MQCL) and INST 95/1654-1 (PQET), Germany's Excellence Strategy (MCQST, EXC-2111, 390814868), the Bavarian State Ministry of Science and Arts via the project EQAP. C.A.-S. acknowledges the support from the Comunidad de Madrid fund “Atracci\'on de Talento, Mod. 1”, Ref. 2020-T1/IND-19785, the projects from the Ministerio de Ciencia e Innovaci\'on PID2023-148061NB-I00 and PCI2024-153425, the project ULTRABRIGHT from the Fundaci\'on Ram \'on Areces, the Grant “Leonardo for researchers in Physics 2023” from Fundaci\'on BBVA, and the Spanish State through the Recovery, Transformation, and Resilience Plan (MAD2D-CM-UAM7), and the European Union through the Next Generation EU funds. F.P.L. acknowledges support from HORIZON EIC-2022PATHFINDERCHALLENGES-01 HEISINGBERG Project 101114978. E.d.V. acknowledges support from the CAM Pricit Plan (Ayudas de Excelencia del Profesorado Universitario), the Technical University of Munich – Institute for Advanced Study (Hans Fischer Fellowship) and the Spanish Ministry of Science, Innovation and Universities through the ``Maria de Maetzu'' Programme for Units of Excellence in R\&D (CEX2023-001316-M), the MCIN/AEI/10.13039/501100011033, FEDER UE, projects No.~PID2020-113415RB-C22 (2DEnLight) and No.~PID2023-150420NB-C31 (Q), and from the Proyecto Sinérgico CAM 2020 Y2020/TCS-6545 (NanoQuCo-CM). S.K.K. and C.A.-S. thank D. Marni-Sobrino for his collaboration in the phase stabilization implementation. \\

\section*{Author contributions}
E.Z.C., F.P.L. and E.d.V. developed the theoretical framework.
S.K.K. and L.H. executed the experiments with contributions from K.B., F.S. and C.C., under supervision of K.M. and J.J.F.
C.A.-S. implemented the phase control and stabilization technique.
S.K.K., E.Z.C. and L.H. were responsible for data analysis. 
H.R. prepared the sample under study.
E.d.V. and K.M. organized the research.
The manuscript was written by S.K.K. and F.P.L. with input from all the authors.\\

\providecommand{\noopsort}[1]{}\providecommand{\singleletter}[1]{#1}%

\newpage

\begin{widetext}
	\makeatletter
	\def\MakeTitle{
		\@author@finish
		\title@column\titleblock@produce
		\suppressfloats[t]}
	\makeatother
	\title{Unlocking multiphoton emission from a single-photon source \\through mean-field engineering
		\\
		Supplementary information}
	\date{\today}

	\MakeTitle
	\onecolumngrid

	\subsection{System characterization: mean field, fluctuations and driving}
	In our experimental configuration depicted in Fig.~\ref{fig1}(e) in the manuscript, the system~$\sigma$---the resonance fluorescence of our TLS---interferes with an external coherent field given by Eq.~(\ref{eq:Mon16Sep120257CEST2024}) in the homodyne beam splitter~(BS). The two outputs of the BS are written as~$s_+ = \left(\sigma+\mathcal{F}\langle\sigma\rangle e^{i\phi}\right)/\sqrt{2}$ and~$s_-=\left(\sigma-\mathcal{F}\langle\sigma\rangle e^{i\phi}\right)/\sqrt{2}$ with variable phase~$\phi$. In the manuscript, the factor of~$1/\sqrt{2}$ is omitted for simplicity. Their intensities are expressed with the TLS operator and the fluctuation operator as 
	$
	\langle {s_\pm}^\dagger s_\pm\rangle =  \frac{1}{2} \left[\langle{\varsigma^\dagger\varsigma}\rangle+|\langle{\sigma}\rangle|^2 \left(1\pm2\mathcal{F}\cos{(\phi)+\mathcal{F}^2}\right)\right]
	$
	, with an interference term varying with phase and the LO amplitude~$\mathcal{F}$. As mentioned in the manuscript, the measured intensities are proportional to the corresponding quantities. We can express the measured intensities,~$I_{s\pm}\propto\langle{s_\pm}^\dagger s_\pm\rangle$, of the two BS outputs as
	\begin{equation}
		I_{s\pm} =  \frac{1}{2} \left[I_\varsigma+I_{\langle{\sigma}\rangle} \left(1\pm2\mathcal{F}\cos{(\phi)+\mathcal{F}^2}\right)\right]\,.
		\label{suppl_eqn_driving_measurement}
	\end{equation}
	One can extract the intensity of the coherent field,~$I_{\langle{\sigma}\rangle}\propto|\langle{\sigma}\rangle|^2$, and that of the fluctuations,~$I_\varsigma\propto\langle{\varsigma^\dagger\varsigma}\rangle$, by examining the~$\mathcal{F}$- and~$\phi$-dependent admixture signal intensities given by Eq.~\ref{suppl_eqn_driving_measurement}. Experimentally, we drive a piezo stage to continuously scan the phase, by varying the path length difference between the LO and the emission of the system, while measuring intensities of the two outputs for a given driving~$\Omega$ and a constant LO intensity~$\propto\mathcal{F}^2$. An exemplary time-traced counter measurement result of the two outputs for a excitation power of~$\SI{2}{\uW}\propto\Omega^2$ and a LO intensity of~$\SI{189.1}{cts/ms}$ is shown in Fig.~\ref{suppl_fig_dirving}(a). In Eq.~(\ref{suppl_eqn_driving_measurement}), the maxima and minima occur with~$\phi=n\pi$, where~$n$ is an integer value. The phase scan with a large range allows us to observe multiple extrema in the results. We evaluate the average maximum and minimum for the given parameters of~$\mathcal{F}$ and the excitation power, by considering extrema in every oscillation period from the whole time-traced measurement. Next, we study the~$\mathcal{F}$-dependency of these averaged maximum and minimum counts for the same driving and present the result in Fig.~\ref{suppl_fig_dirving}(b). The minima and maxima are given by red and blue symbols, corresponding to the positive and negative sign in Eq.~(\ref{suppl_eqn_driving_measurement}) with~$\phi=\pi$. The equation is used as a fitting function with two fitting parameters~$I_\varsigma$ and~$I_{\langle{\sigma}\rangle}$. As a result, we obtain the intensity of the coherent mean field of the system~$I_{\langle{\sigma}\rangle}\approx\SI{252.2}{cts/\ms}$ and that of the fluctuations~$I_\varsigma\approx\SI{47.5}{cts/\ms}$. An excellent agreement between measured data and solid fitting curves is found in the figure. The driving~$\Omega=\sqrt{(\langle{\varsigma^\dagger\varsigma}\rangle)/(8|\langle{\sigma}\rangle|^2)}$ can be derived from Eq.~(\ref{eqn_intensities}) in the main manuscript. Given the intensities of the two components, a driving~$\Omega\approx0.15$ is estimated. Similarly, we characterize the system with two different drivings used in the power series of~Fig.~\ref{fig4}. As a result, $I_{\langle{\sigma}\rangle}\approx\SI{849.7}{cts/\ms}$ and~$I_\varsigma\approx\SI{521.4}{cts/\ms}$ for~$\Omega\approx0.28$ as well as~$I_{\langle{\sigma}\rangle}\approx\SI{1193.1}{cts/\ms}$ and~$I_\varsigma\approx\SI{1504.0}{cts/\ms}$ for~$\Omega\approx0.40$ are extracted.
	
	\begin{figure*}[!h]
		\centering
		\includegraphics[width=\textwidth]{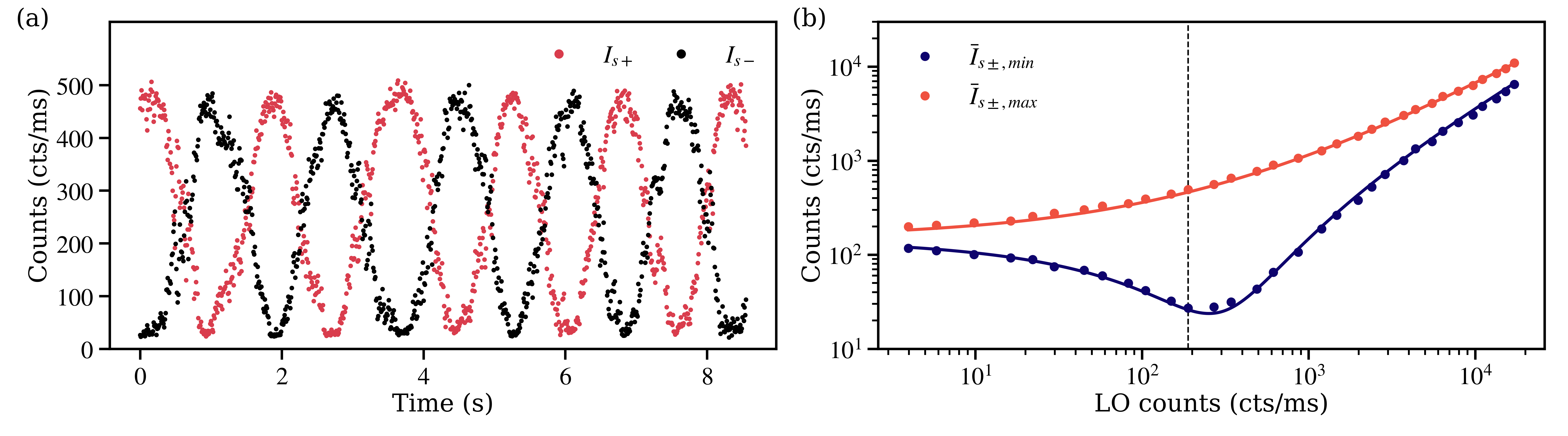}
		\caption{(a) Time-traced counts of the two homodyne BS outputs,~$I_{s+}$ and~$I_{s-}$, while scanning phase~$\phi$ continuously with a constant excitation power of~$\SI{2}{\uW}$ and a LO intensity of~$\SI{189.1}{cts/ms}$. (b) The averaged minima~$\bar{I}_{s\pm,\text{min}}$ and maxima~$\bar{I}_{s\pm,\text{max}}$ from phase-scanned measurements with varying LO intensity are shown as blue and red symbols. By fitting the theoretical model~(solid lines) to the experimental data, the intensities of the mean field~$I_{\langle{\sigma}\rangle}\approx\SI{252.2}{cts/\ms}$ and fluctuations~$I_\varsigma\approx\SI{47.5}{cts/\ms}$ as well as the driving~$\Omega\approx0.15$ are obtained. The dashed line represents the LO intensity, where the results in (a) are obtained.}
		\label{suppl_fig_dirving}
	\end{figure*}
	
	\subsection{Analysis of multiphoton correlations}
	To study the three-photon correlation, we extend the standard Hanbury Brown-Twiss setup with two outputs to the modified version with three outputs as shown in Fig.~\ref{fig1}(e). As mentioned in the manuscript, we characterize~$G^{(3)}(\tau_1,\tau_2)$ based on the time difference between photon arrival times at the three detectors. To extract~$g^{(3)}(\tau^{*})$ from the time correlation measurement~$G^{(3)}(\tau_1,\tau_2)$, we perform an analysis which projects the two-dimensional data~($\tau_1,\tau_2$-dependent) onto the one-dimensional space~($\tau^*$-dependent). First, the obtained data are transformed from the Cartesian coordinate system with~$(\tau_1,\tau_2)$ to a polar coordinate system with~$(\tau^*, \theta)$. 
	As an example, the~$G^{(3)}(\tau_1, \tau_2)$ result for a driving~$\Omega\approx0.15$ without the external LO and its transformed data~$G^{(3)}(\tau^*, \theta)$ are shown in Fig.~\ref{suppl_fig_g3int}(a) and (b), respectively. 
	For the transformation, we define~$\tau^*=\tau_1/\cos{\theta}$, where~$\theta$ is the angle with respect to the~$\tau_1$ axis in the Cartesian coordinate system. The transformation allows us to integrate the data along the~$\theta$ axis, preserving the time parameter~$\tau^*$ which effectively contains time delay information of both~$\tau_1$ and~$\tau_2$. Consequently, we can evaluate~$G^{(3)}(\tau^*)$, where~$\tau^*$ can be interpreted as the effective time delay of three subsequent detection events by all three detectors. Specifically, we define it as
	\begin{equation}
		G^{(3)}(\tau^*) = \int_{\vartheta}^{\pi/2-\vartheta} G^{(3)}(\tau^*, \theta)~d\theta\,,
		\label{suppl_eqn_G3_tau}
	\end{equation}
	where the integration range~$(\vartheta, \pi/2-\vartheta)$ is centered at~$\pi/4$. The integration center angle~$\pi/4$ corresponds to the antidiagonal line~($\tau_1 = -\tau_2$) in the Cartisean coordinate system, where the most uncorrelated events occur due to the largest time difference between~$\tau_1$ and~$\tau_2$ for a given~$\tau^*$. We carefully choose~$\vartheta =\pi/12$, making the integration range wide enough to benefit from a large integration window, improvement of signal-to-noise ratio, while still narrow enough to exclude the three diagonals from the integration, namely,~$G^{(3)}(0,\tau_2)$, $G^{(3)}(\tau_1,0)$ and $G^{(3)}(\tau_1,\tau_1)$ lines representing two-photon coincidences. The normalized correlation is given by~$g^{(3)}(\tau^*)=G^{(3)}(\tau^*) / \bar{G}^{(3)}_{\infty}$, where~$\bar{G}^{(3)}_{\infty}$ is the averaged uncorrelated events at~$|\tau^*|\gg0$ which are positioned at both ends of~$G^{(3)}(\tau^*)$.

	For the second-order correlation~$g^{(2)}(\tau)$, we first extract the unnormalized~$G^{(2)}(\tau)$ by analyzing the same raw time-tag data which are used for the~$g^{(3)}(\tau^*)$ analysis. To improve and balance counts, we combine the two channels out of the three of the extended Hanbury Brown-Twiss setup which are passing two BSs in Fig.~\ref{fig1}(e). A cross correlation analysis is performed on the combined channel with the other to obtain~$G^{(2)}(\tau)$ . Similar to the normalization step for~$g^{(3)}(\tau^*)$, we evaluate~$g^{(2)}(\tau)$ by dividing~$G^{(2)}(\tau)$ by~$\bar{G}^{(2)}_{\infty}$. The standard deviation in the averaged uncorrelated events~$\bar{G}^{(n)}_{\infty}$ and the Poissonian statistics of~$G^{(n)}(0)$ are taken into account to calculate errors of~$g^{(n)}(0)$ for~$n=2, 3$.
	\begin{figure*}[!h]
		\centering
		\includegraphics[width=\textwidth]{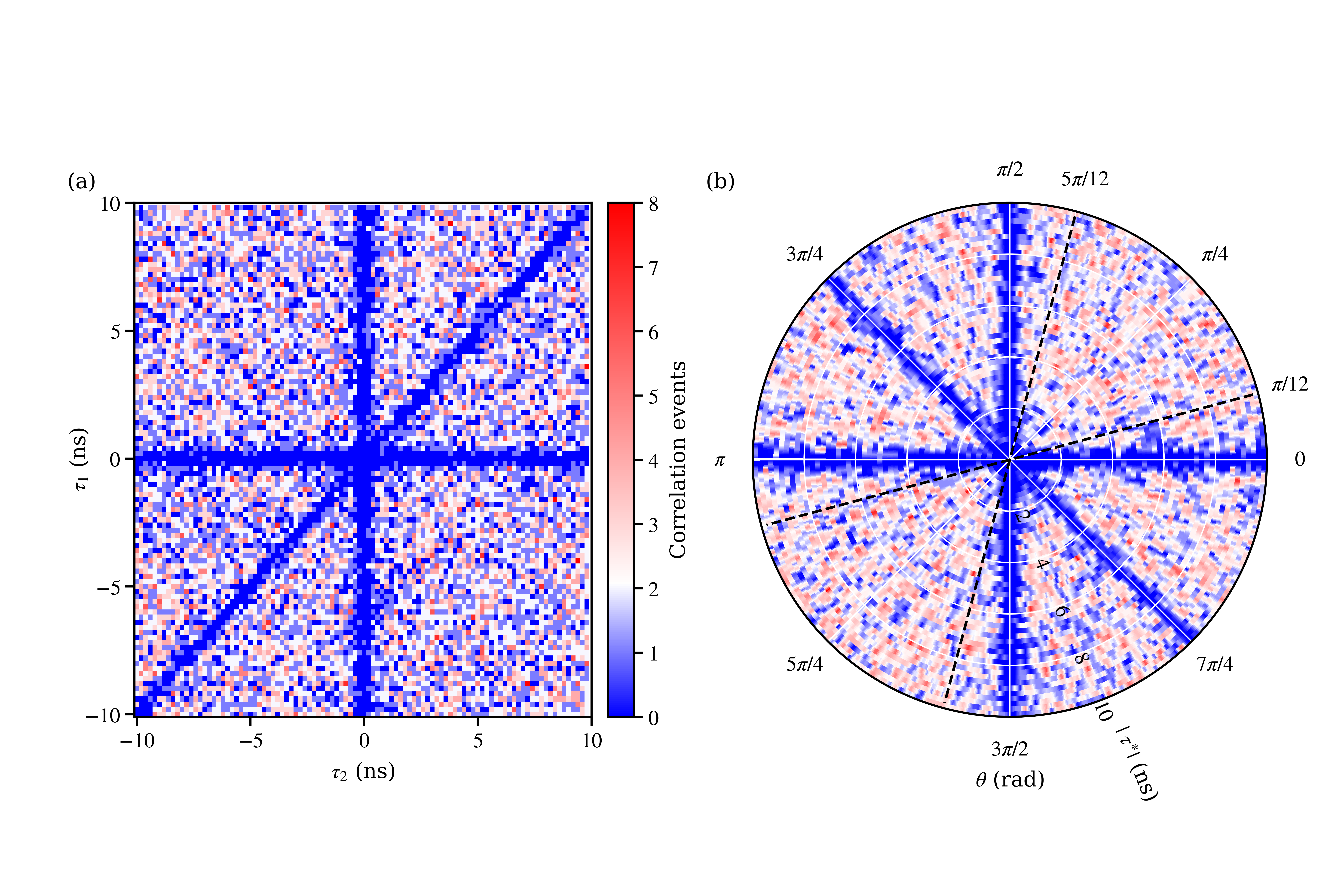}
		\caption{(a) With time difference of three detector channels, unnormalized third-order correlation~$G^{(3)}(\tau_1, \tau_2)$ is obtained in the Cartesian coordinate system under the measurement condition of~$\Omega\approx0.15$ and~$\mathcal{F} =0$. (b) The result is transformed in a polar coordinate system as~$G^{(3)}(\tau^*, \theta)$. We perform the radial integration along the~$\theta$ axis with the integration window centered~$\phi=\pi/4$. Dashed lines represent the limit of integration angles~$\left(\pi/12\right)$ and~$\left(\pi/2-\pi/12\right)$.}
		\label{suppl_fig_g3int}
	\end{figure*}

	\providecommand{\noopsort}[1]{}\providecommand{\singleletter}[1]{#1}%

\end{widetext}

\end{document}